\documentclass[12pt]{article}
\usepackage{amsmath}
\usepackage{graphicx}
\usepackage{bm}
\usepackage{fancyhdr}
\usepackage{amssymb}
\usepackage{setspace}
\usepackage{epsfig}
\usepackage{overpic}
\usepackage{caption,subcaption}

\begin{document}

%************************** Text Begins here ******************************

%  Greek letters

\def\a{\alpha}
\def\b{\beta}
\def\d{\delta}
\def\e{\epsilon}
\def\g{\gamma}
\def\h{\mathfrak{h}}
\def\k{\kappa}
\def\l{\lambda}
\def\o{\omega}
\def\p{\wp}
\def\r{\rho}
\def\t{\tau}
\def\s{\sigma}
\def\z{\zeta}
\def\x{\xi}
\def\V={{{\bf\rm{V}}}}
 \def\A{{\cal{A}}}
 \def\B{{\cal{B}}}
 \def\C{{\cal{C}}}
 \def\D{{\cal{D}}}
\def\K{{\cal{K}}}
\def\O{\Omega}
\def\R{\bar{R}}
\def\T{{\cal{T}}}
\def\L{\Lambda}
\def\f{E_{\tau,\eta}(sl_2)}
\def\E{E_{\tau,\eta}(sl_n)}
\def\Zb{\mathbb{Z}}
\def\Cb{\mathbb{C}}

\def\R{\overline{R}}
% Shorthands for \begin{equation} and the like

\def\beq{\begin{equation}}
\def\eeq{\end{equation}}
\def\bea{\begin{eqnarray}}
\def\eea{\end{eqnarray}}
\def\ba{\begin{array}}
\def\ea{\end{array}}
\def\no{\nonumber}
\def\le{\langle}
\def\re{\rangle}
\def\lt{\left}
\def\rt{\right}

\newtheorem{Theorem}{Theorem}
\newtheorem{Definition}{Definition}
\newtheorem{Proposition}{Proposition}
\newtheorem{Lemma}{Lemma}
\newtheorem{Corollary}{Corollary}
\newcommand{\proof}[1]{{\bf Proof. }
        #1\begin{flushright}$\Box$\end{flushright}}

\baselineskip=20pt

%%%%%%%%%%%%%%%%%%%%%%%%%%%%%%%%%%%%%%%%%%%%%%%%%%%%%%%%%%%%
%                                                          %
%  Title page                                              %
%                                                          %
%%%%%%%%%%%%%%%%%%%%%%%%%%%%%%%%%%%%%%%%%%%%%%%%%%%%%%%%%%%%
\newfont{\elevenmib}{cmmib10 scaled\magstep1}
\newcommand{\preprint}{
   \begin{flushleft}
     %\elevenmib Yukawa\, Institute\, Kyoto\\
   \end{flushleft}\vspace{-1.3cm}
   \begin{flushright}\normalsize
  % \sf  YITP-03-53\\
   %  {\tt hep-th/yymmnnn} \\ November 2005
   \end{flushright}}
\newcommand{\Title}[1]{{\baselineskip=26pt
   \begin{center} \Large \bf #1 \\ \ \\ \end{center}}}
\newcommand{\Author}{\begin{center}
   \large \bf
Yi Qiao${}^{a,b,c}$, Jian Wang${}^{b,d}$, Junpeng Cao${}^{b,d,e}\footnote{Corresponding author: junpengcao@iphy.ac.cn}$ and Wen-Li Yang${}^{a,c,f}\footnote{Corresponding author:
wlyang@nwu.edu.cn}$
 \end{center}}
\newcommand{\Address}{\begin{center}
     ${}^a$ Institute of Modern Physics, Northwest University,
     Xian 710127, China\\
     ${}^b$ Beijing National Laboratory for Condensed Matter
           Physics, Institute of Physics, Chinese Academy of Sciences, Beijing
           100190, China\\
     ${}^c$ Shaanxi Key Laboratory for Theoretical Physics Frontiers,  Xian 710127, China\\
     ${}^d$ School of Physical Sciences, University of Chinese Academy of Sciences, Beijing, China \\
     ${}^e$ Songshan Lake Materials Laboratory, Dongguan, Guangdong 523808, China\\
     ${}^f$ School of Physics, Northwest University, Xian 710127, China
   \end{center}}

\preprint \thispagestyle{empty}
\bigskip\bigskip\bigskip

\Title{Exact solution of a anisotropic $J_1-J_2$ spin chain with antiperiodic boundary condition} \Author

\Address \vspace{0.1cm}
\begin{abstract}
The exact solution of an integrable anisotropic Heisenberg spin chain with nearest-neighbour, next-nearest-neighbour and scalar chirality couplings is studied, where the
boundary condition is the antiperiodic one. The detailed construction of Hamiltonian and the proof of integrability are given.
The antiperiodic boundary condition breaks the $U(1)$-symmetry of the system and we use the off-diagonal Bethe Ansatz to solve it.
The energy spectrum is characterized by the inhomogeneous $T-Q$ relations and the contribution of the inhomogeneous term is studied.
The ground state energy and the twisted boundary energy in different regions are obtained.
We also find that the Bethe roots at the ground state form the string structure if the coupling constant $J=-1$ although the Bethe Ansatz equations are the inhomogeneous ones.

\vspace{1truecm}
%75.10.Pq Spin chain models
%02.30.Ik Integrable systems
%71.10.Pm Fermions in reduced dimensions (anyons, composite fermions, Luttinger liquid, etc.

\noindent {\it PACS:} 75.10.Pq, 03.65.Vf, 71.10.Pm

\noindent {\it Keywords:} Quantum spin chain; Bethe Ansatz; Yang-Baxter equation.
\end{abstract}

\newpage

%\tableofcontents

%%%%%%%%%%%%%%%%%%%%%%%%%%%%%%%%%%%%%%%%%%%%%%%%%%%%%%%%%%%%%%%
%                                                             %
%  1. Introduction                                            %
%                                                             %
%%%%%%%%%%%%%%%%%%%%%%%%%%%%%%%%%%%%%%%%%%%%%%%%%%%%%%%%%%%%%%%
\hbadness=10000

\tolerance=10000

\hfuzz=150pt

\vfuzz=150pt

\section{Introduction}\label{intro}
\setcounter{equation}{0}

The Heisenberg model is a typical system to describe the quantum magnetism, where the spin exchanging interaction is the nearest-neighbor (NN) one. A nontrivial generalization
of the Heisenberg model is the $J_1-J_2$ model, where the NN and the next-nearest-neighbor (NNN) interactions are involved \cite{93,97,98,Sha81,Nom92,Jaf06,Djo16}.
Many interesting phenomena have been found in the $J_1-J_2$ model. For example,
at the point of $J_2/J_1=0.241$, the $J_1-J_2$ model has a topological phase transition \cite{95,Jaf07}.
At the Majumdar-Ghosh point, $J_2/J_1=0.5$, the model Hamiltonian degenerates into a projector operator and only the ground state can be obtained exactly \cite{92}.

Although the $J_1-J_2$ model can not be solved exactly, people find that the $J_1-J_2$ model with some additional terms is integrable.
For example, Popkov and Zvyagin proposed the integrable two-chain and multichain quantum spin model \cite{Pop93,Zvy95-1,Zvy95-2,Zvy01}.
Frahm and R\"odenbeck constructed an integrable model of two coupled Heisenberg chains by taking the derivative of the logarithm of product of two transfer matrices with different spectral parameters \cite{Fra96,Fra97,Fra14}.
Using the samilar idea, Ikhlef, Jacobsen and Saleur constructed the $Z_2$ staggered vertex model \cite{Ikh08,Ikh10}.
These models are equivalent to the $J_1-J_2$ model with some spin chirality terms, where the extra scalar chirality terms are introduced to ensure the integrability.
Tavares and Ribeiro studied the thermodynamic properties of this kind of models by using the quantum transfer matrix method \cite{Tav13,Tav14}.
Recently, the models with chirality terms have attracted lot of interest in the context of quantum spin liquids \cite{Gor15,Che17}.

The quantization condition used in the above references are the periodic boundary condition. In this paper,
we study the integrable anisotropic $J_1-J_2$ spin chain with antiperiodic boundary condition. We note that in this case, the
$U(1)$ symmetry of the system is broken and the traditional Bethe ansatz method does not work due to the lack of reference state.
The antiperiodic (twisted) boundary condition is tightly related to the recent study on the topological states of matter.
The model Hamiltonian considered in this paper is
\bea\label{Ham}
H &=&J\sum^{2N}_{j=1} \bigg\{ \cosh(2a)(\sigma_j^x \sigma_{j+1}^x+\sigma_j^y \sigma_{j+1}^y)+\cosh\eta \sigma_j^z \sigma_{j+1}^z \no \\
&&-\frac{\sinh^2(2a)\cosh\eta}{2\sinh^2\eta}\vec{\sigma}_j \cdot \vec{\sigma}_{j+2}+\frac{(-1)^j i \sinh(2a)}{2 \sinh\eta} \big\{ \cosh\eta \vec{\sigma}_{j+1} \!\cdot\!(\vec{\sigma}_{j}  \times \vec{\sigma}_{j+2} ) \no \\
&&+[\cosh(2a)\!-\!\cosh\eta] \sigma_{j+1}^z(\sigma_{j}^x\sigma_{j+2}^y \!-\! \sigma_{j}^y\sigma_{j+2}^x) \big\}\! \bigg\},
\eea
where $\vec{\sigma}_j\equiv (\sigma^x_j,\, \sigma^y_j,\, \sigma^z_j)$ are the Pauli matrices at site $j$, $a$ and $\eta$ are the generic constants describing the coupling strengths,
and the boundary condition is the antiperiodic one
\bea
\sigma^{\alpha}_{2N+n}=\sigma^{x}_{n} \sigma^{\alpha}_{n} \sigma^{x}_{n},\quad  n=1,2, \quad \alpha=x,y,z.\label{Periodic-C}
\eea
In the Hamiltonian (\ref{Ham}), the first two terms describe an anisotropic NN interaction, the third term is an isotropic NNN interaction and the last one corresponds to an anisotropic chiral three-spin interaction.
We note that the hermitian of the Hamiltonian (\ref{Ham}) requires that $a$ must be real if $\eta$ is imaginary (gapped regime), and $a$ must be imaginary if $\eta$ is real (gapless regime).
We use the off-diagonal Bethe Ansatz (ODBA) \cite{Cao13,Wan15} to solve the model.

The paper is organized as follows. In the next section, we prove that the model (\ref{Ham}) is integrable. In section \ref{3}, we derive the exact energy spectrum and the Bethe Ansatz equations. Ground state and twisted boundary energy with $J=1$ are given in section \ref{4} and the corresponding results with $J=-1$ are discussed in section \ref{5}. Section \ref{6} is attributed to the concluding remarks.

%%%%%%%%%%%%%%%%%%%%%%%%%%%%%%%%%%%%%%%%%%%%%%%%%%%%%%%%%%%%%%%
%                                                             %
%   NNA torus model                                           %
%                                                             %
%                                                             %
%                                                             %
%%%%%%%%%%%%%%%%%%%%%%%%%%%%%%%%%%%%%%%%%%%%%%%%%%%%%%%%%%%%%%%

\section{Integrability}\label{2}
\setcounter{equation}{0}

Throughout, ${V}$ denotes a two-dimensional linear space and let $\{|m\rangle, m=0,1\}$ be an orthogonal basis of it.
We shall adopt the standard notations: for any matrix $A\in {\rm End}({ V})$, $A_j$ is an
embedding operator in the tensor space ${ V}\otimes
{ V}\otimes\cdots$, which acts as $A$ on the $j$-th space and as
identity on the other factor spaces. For $B\in {\rm End}({ V}\otimes { V})$, $B_{ij}$ is an embedding
operator of $B$ in the tensor space, which acts as identity
on the factor spaces except for the $i$-th and $j$-th ones.

Let us introduce the $R$-matrix $R_{0,j}(u)\in {\rm End}({ V}_0\otimes { V}_j)$
\begin{eqnarray}\label{R-matrix}
R_{0,j}(u) \!&=&\! \frac{1}{2} \bigg[ \frac{\sinh(u\!+\!\eta)}{\sinh \eta} (1\!+\!\sigma^z_0 \sigma^z_j) \!+\! \frac{\sinh u}{\sinh \eta} (1\!-\!\sigma^z_0 \sigma^z_j)
 \bigg]  + \frac{1}{2} (\sigma^x_0 \sigma^x_j +\sigma^y_0 \sigma^y_j)
\no \\ \!&=&\!\frac{1}{\sinh\eta} \! \left(\! \begin{array}{cccc}
\sinh(u+\eta) & 0    & 0    & 0 \\
0      & \sinh u    & \sinh\eta & 0 \\
0      & \sinh\eta & \sinh u    & 0 \\
0      & 0    & 0    & \sinh(u+\eta)
\end{array} \!\right)\!,
\end{eqnarray}
where $u$ is the spectral parameter.
The $R$-matrix (\ref{R-matrix}) satisfies the following relations
\bea
&&\hspace{-0.5cm}\mbox{ Initial
condition}:\,R_{0,j}(0)=  P_{0,j},\no \\
&&\hspace{-0.5cm}\mbox{ Unitary
relation}:\,R_{0,j}(u)R_{j,0}(-u)= \phi(u)\times {\rm id},\no \\
&&\hspace{-0.5cm}\mbox{ Crossing
relation}:\,R_{0,j}(u)=-\sigma_0^y R_{0,j}^{t_0}(-u-\eta)\sigma_0^y,\no \\
&&\hspace{-0.5cm}\mbox{ PT-symmetry}:\,R_{0,j}(u)=R_{j,0}(u)=R_{0,j}^{t_0\,t_j}(u),\label{PT}
\eea
where $\phi(u)=-\sinh(u+\eta)\sinh(u-\eta)/\sinh^2\eta$, $t_0$ (or $t_j$) denotes the
transposition in the space ${V}_0$ (or ${V}_j$) and
$P_{0,j}$ is the permutation operator possessing the property
\begin{eqnarray}
  R_{j,k}(u)=P_{0,j}R_{0,k}(u)P_{0,j}.
\end{eqnarray}
The $R$-matrix (\ref{R-matrix}) satisfies the Yang-Baxter equation (YBE)
\bea
&&R_{1,2}(u_1-u_2)R_{1,3}(u_1-u_3)R_{2,3}(u_2-u_3)
\no \\ && \quad =R_{2,3}(u_2-u_3)R_{1,3}(u_1-u_3)R_{1,2}(u_1-u_2).\label{QYB}
\eea

We define the monodromy matrices as
\bea\label{monodromy-matrix}
T_0(u)&=&\sigma^x_0 R_{0,1}(u+a) R_{0,2}(u-a) \cdots R_{0,2N-1}(u+a) R_{0,2N}(u-a), \no \\
\hat{T}_0(u)&=&\sigma^x_0 R_{0,2N}(u+a) R_{0,2N-1}(u-a) \cdots R_{0,2}(u+a) R_{0,1}(u-a),
\eea
where $V_0$ is the auxiliary space, $V_1\otimes V_2 \otimes \cdots \otimes V_{2N}$ is the physical or quantum space,
$2N$ is the number of sites and $a$ is the inhomogeneous parameter.
From the YBE (\ref{QYB}) and the fact
\bea
[R_{1,2}(u), \sigma^x_1\sigma^x_2]=0,
\eea
one can prove that the monodromy matrix $T(u)$ satisfies the Yang-Baxter relation
\bea
R_{1,2}(u-v)T_1(u)T_2(v)= T_2(v)T_1(u)R_{1,2}(u-v).\label{RTT}
\eea
The transfer matrices are the trace of monodromy matrices in the auxiliary space
\bea \label{trans}
t(u)=tr_0 T_0(u), \quad  \hat{t}(u)=tr_0 \hat{T}_0(u).
\eea
Using the crossing symmetry in Eq.(\ref{PT}), we obtain the relations between transfer matrices $t(u)$ and $\hat t(u)$
\bea\label{tt}
t(u)=-\hat{t}(-u-\eta), \quad \hat{t}(u)=-t(-u-\eta).
\eea
From the Yang-Baxter relation (\ref{RTT}) and Eq.(\ref{tt}), one can prove that the transfer matrices
$t(u)$ [or $\hat t(u)$] with different spectral parameters
commute with each other. Meanwhile, the transfer matrices $t(u)$ and $\hat t(u)$ also commute with each other
\bea
[t(u), t(v)]=[\hat t(u), \hat t(v)] =[t(u), \hat t(u)]=0. \label{t-commu1} \eea
Therefore, both $t(u)$ and $\hat t(u)$ serve as the generating functions of all the conserved quantities of the
system, and the transfer matrices $t(u)$ and $\hat t(u)$ can be diagonalized simultaneously.

Using the initial condition of the $R$-matrix (\ref{PT}), we obtain
\bea
&&\hat{t}(-a) =      R_{2N,2N-1}(-2a) \cdots  R_{2N,2}(0) R_{2N,1}(-2a) \sigma^{x}_{2N}, \no \\[4pt]
&&\hat{t}(a) =   \sigma^{x}_{1} R_{1,2N}(2a) R_{1,2N-1}(0) \cdots  R_{1,2}(2a).\label{t1}
\eea
Taking the derivative of transfer matrix $t(u)$ with respect to $u$ and consider the values at the point of $u=a$, we have
\bea\label{t3}
&&\frac{\partial \, t(u)}{\partial u}\big|_{u=a}
= \sum^{N-1}_{j=1} \sigma^{x}_{2N} R_{2N,1}(2a) R_{2N,2}(0) [ R'_{2N,2j-1}(2a) R_{2N,2j}(0) + R_{2N,2j-1}(2a)
\no \\[4pt] && \quad \times R'_{2N,2j}(0) ] \cdots R_{2N,2N-1}(2a)
+ \sigma^{x}_{2N} R_{2N,1}(2a) R_{2N,2}(0) \cdots R'_{2N,2N-1}(2a)
\no \\[4pt] && \quad \times + R_{2,3}(2a)
R_{2,4}(0) \cdots R_{2,2N-1}(2a) R'_{2,2N}(0)\sigma^{x}_{2}R_{2,1}(2a),
\eea%\no \\[4pt] && \quad \times
where $R'_{i,j}(u)=\frac{\partial }{\partial u}\,R_{i,j}(u)$. Similarly the derivative of $t(u)$ at the point of $u=-a$ is
\bea\label{t4}
&& \frac{\partial \, t(u)}{\partial u}\big|_{u=-a} = \sum^{N}_{j=1} R_{1,2}(-2a) \cdots [ R'_{1,2j-1}(0) R_{1,2j}(-2a) + R_{1,2j-1}(0) R'_{1,2j}(-2a) ] \cdots
\no \\[4pt] && \quad \times R_{1,2N-1}(0) R_{1,2N}(-2a)\sigma^{x}_{2}
+ R_{1,2}'(-2a) R_{1,3}(0) \cdots R_{1,2N}(-2a)\sigma^{x}_{1}
\no \\[4pt] && \quad \times + R_{2N-1,2N}(-2a)\sigma^{x}_{2N-1}R_{2N-1,1}(0)R_{2N-1,2}(-2a) \cdots R_{2N-1,2N-2}(-2a).
\eea

The integrable Hamiltonian can be constructed from the transfer matrices $t(u)$ and $\hat t(u)$ as
\bea
H &=& \frac{JN\cosh\eta[\cosh^2(2a)-\cosh(2\eta)]}{\sinh^2\eta} +J\phi^{1-N}(2a)\sinh\eta
\no \\ && \times J\bigg\{ \hat{t}(-a)\frac{\partial \, t(u)}{\partial u}\big|_{u=a}+ \hat{t}(a) \frac{\partial \, t(u)}{\partial u}\big|_{u=-a} \bigg\}. \label{t-c2ommu1}
\eea
Substituting the relations (\ref{t1})-(\ref{t4}) into above expression (\ref{t-c2ommu1}), we obtain
\bea\label{Ham-1}
H&=& \sinh\eta \bigg\{ \sum^{N-1}_{j=1}
\big[  R_{2j,2j-1}(-2a) R'_{2j,2j-1}(2a) + R_{2j+1,2j}(2a) R'_{2j+1,2j}(-2a)
\no \\[4pt] && + R_{2j+2,2j+1}(-2a) P_{2j+2,2j}R'_{2j+2,2j}(0) R_{2j+2,2j+1}(2a)+R_{2j+1,2j}(2a)
\no \\[4pt] && \times P_{2j+1,2j-1} R'_{2j+1,2j-1}(0) R_{2j+1,2j}(-2a) \big]
\no \\[4pt] && + R_{2N,2N-1}(-2a) R'_{2N,2N-1}(2a) + \sigma^x_1 R_{1,2N}(2a) R'_{2j+1,2j} (-2a)\sigma^x_1 \no \\[4pt] && + R_{2,1}(-2a) \sigma^x_2 P_{2,2N}R'_{2,2N}(0) \sigma^x_2 R_{2,1}(2a)
+\sigma^x_1 R_{1,2N}(2a) \no \\[4pt] && \times P_{1,2N-1} R'_{1,2N-1}(0) R_{1,2N}(-2a)\sigma^x_1
\bigg\} -\frac{N\cosh\eta[\cosh^2(2a)-\cosh(2\eta)]}{\sinh^2\eta}.
\eea
The derivative of the $R$-matrix reads
\bea \label{R'-matrix}
R'_{0,j}(u)=\frac{1}{2}\bigg[ \frac{\cosh(u+\eta)}{\sinh\eta}(1+\sigma_0^z\sigma_j^z) + \frac{\cosh u}{\sinh\eta} (1-\sigma_0^z\sigma_j^z) \bigg].
\eea
The commutative relation between the permutation operators is
\bea
[P_{2,1},P_{2,0}]
&=&\frac{1}{4}[(1+{\vec\sigma}_{2}\cdot{\vec\sigma}_{1}),(1+{\vec\sigma}_{2}\cdot{\vec\sigma}_{0})] \no \\[4pt]
&=& \frac{i}{2} {\vec\sigma}_2\cdot ({\vec\sigma}_1 \times {\vec\sigma}_0).\label{Rx}
\eea
Substituting Eqs.(\ref{R'-matrix}) and (\ref{Rx}) into (\ref{Ham-1}) and after some tedious calculations, we arrive at the Hamiltonian (\ref{Ham}).
From the construction (\ref{t-c2ommu1}) and the commutation relation (\ref{t-commu1}) of generating functions $t(u)$ and $\hat t(u)$, we conclude that
the model (\ref{Ham}) with the antiperiodic boundary condition is integrable.

\section{Exact solution}\label{3}
\setcounter{equation}{0}

We first introduce the inhomogeneous monodromy matrix
\bea\label{monodromy-matrix-1}
T^g_0(u)=\sigma^x_0 R_{0,1}(u-\theta_1) R_{0,2}(u-\theta_2) \cdots R_{0,2N-1}(u-\theta_{2N-1}) R_{0,2N}(u-\theta_{2N}),
\eea
where the $\{\theta_j, j=1,\cdots, N\}$ are the inhomogeneous parameters.
The matrix form of monodromy matrix $T^g_0(u)$ in the auxiliary space is
\begin{equation}\label{monodromy-matrix1}
  T_0^g(u)=\left(
             \begin{array}{cc}
               C(u) & D(u)\\
               A(u) & B(u)\\
             \end{array}
           \right).
\end{equation}
where $A(u)$, $B(u)$, $C(u)$ and $D(u)$ are the operators acting in the quantum space.
We denote the all spins aligning up state as the vacuum state $ |0\rangle$,
\bea
|0\rangle=\left(\begin{array}{c}
    1 \\
    0 \\
  \end{array}\right)_1 \otimes\cdots\otimes\left(\begin{array}{c}
    1 \\
    0 \\
  \end{array}\right)_{2N}.\label{left-vacuum}
\eea
The elements of the monodromy matrix $T^g_0(u)$ acting on the vacuum state gives
\bea
&& A(u)|0\rangle=\tilde{a}(u)|0\rangle, \quad B(u)|0\rangle\neq 0,
\no \\ && C(u)|0\rangle=0, \quad \qquad D(u)|0\rangle=\tilde{d}(u)|0\rangle,\label{right-action-1}
\eea
where
\bea\label{aw-dw-fun}
\tilde{a}(u)=\prod_{j=1}^{2N}\frac{\sinh(u-\theta_j+\eta)}{\sinh\eta},\quad
\tilde{d}(u)=\prod_{j=1}^{2N}\frac{\sinh(u-\theta_j)}{\sinh\eta}.\no
\eea
The transfer matrix $t(u)$ defined as
\bea\label{trans1}
t^g(u)=tr_0 T^g_0(u)=B(u)+C(u).
\eea
Suppose $|\Phi\rangle$ is the eigenstate of the transfer matrix $t^g(u)$ and the corresponding eigenvalue is $\Lambda^g(u)$,
\bea
t^g(u)|\Phi\rangle=\Lambda^g(u)|\Phi\rangle.
\eea
According to the results given in \cite{Wan15}, we known that $\Lambda^g(u)$ satisfies following functional relations
\bea
\Lambda^g(\theta_j)\Lambda^g(\theta_j-\eta)=-\tilde{a}(\theta_j)\tilde{d}(\theta_j-\eta),\quad j=1,\cdots,N. \label{ron-1}
\eea
Meanwhile, $\Lambda^g(u)$ is a polynomial of $e^u$ with the degree $2N - 1$ and satisfies the periodicity property
\bea
\Lambda^g(u + i\pi) = (-1)^{2N-1}\Lambda^g(u). \label{ron-2}
\eea
The constraints (\ref{ron-1}) and (\ref{ron-2}) show that the eigenvalue $\Lambda^g(u)$ can be parameterized as the following inhomogeneous $T-Q$ relation
\bea\label{TQ-g}
\Lambda^g(u)=e^u\tilde{a}(u)\frac{Q(u-\eta)}{Q(u)}-e^{-u-\eta}\tilde{d}(u)\frac{Q(u+\eta)}{Q(u)}
-\frac{\tilde{c}(u)\tilde{a}(u)\tilde{d}(u)}{Q(u)},
\eea
where $Q(u)$ is a trigonometric polynomial of the type
\bea
Q(u)=\prod_{j=1}^{2N}\frac{\sinh(u-\lambda_j)}{\sinh\eta}, \label{wworBAEs-inh}
\eea
and $\tilde{c}(u)$ is given by
\bea
\tilde{c}(u)=e^{u-2N\eta+\sum_{l=1}^{2N}(\theta_l-\lambda_l)}-e^{-u-\eta-\sum_{l=1}^{2N} (\theta_l-\lambda_l)}.
\eea
The singularity of $\Lambda^g(u)$ requires that the Bethe roots $\{\lambda_j\}$ in Eq.(\ref{TQ-g}) should satisfy the Bethe ansatz equations (BAEs)
\bea\label{orBAEs-inh}
&&e^{\lambda_j}\tilde{a}(\lambda_j)Q(\lambda_j-\eta)-e^{-\lambda_j-\eta}\tilde{d}(\lambda_j)Q(\lambda_j+\eta)
-\tilde{c}(\lambda_j)\tilde{a}(\lambda_j)\tilde{d}(\lambda_j)=0, \no \\[4pt]
&&\quad j=1,\cdots,N.
\eea

Put $\theta_{2j-1}=-a$ and $\theta_{2j}=a$ for $j=1,\cdots, N$, we obtain the eigenvalue $\Lambda(u)$ of the transfer matrix $t(u)$
\bea\label{TQ}
\Lambda(u)=e^ua(u)\frac{Q(u-\eta)}{Q(u)}-e^{-u-\eta}d(u)\frac{Q(u+\eta)}{Q(u)}
-\frac{c(u)a(u)d(u)}{Q(u)},
\eea
where
\bea\label{adfun}
&&a(u)=\frac{\sinh^N(u+a+\eta)\sinh^N(u-a+\eta)}{\sinh^{2N}\eta}, \no \\
&&d(u)=\frac{\sinh^N(u+a)\sinh^N(u-a)}{\sinh^{2N}\eta},\no\\
&&c(u)=e^{u-2N\eta-\sum_{l=1}^{2N}\lambda_l}-e^{-u-\eta+\sum_{l=1}^{2N}\lambda_l},
\eea
and the BAEs read
\bea\label{orBAEs}
&&e^{\lambda_j}a(\lambda_j)Q(\lambda_j-\eta)-e^{-\lambda_j-\eta}d(\lambda_j)Q(\lambda_j+\eta)
-c(\lambda_j)a(\lambda_j)d(\lambda_j)=0, \no \\[4pt]
&&\quad j=1,\cdots,N.
\eea
From Eqs.(\ref{tt}), (\ref{t-c2ommu1}) and (\ref{TQ}), the energy spectrum of the Hamiltonian (\ref{Ham}) is
\bea\label{energy_real}
E&=& \sum_{j=1}^{2N}J\sinh\eta \, \phi(2a)\bigg\{ \coth(\lambda_j-a)-\coth(\lambda_j-a+\eta)+\coth(\lambda_j+a) \no \\ &-&\coth(\lambda_j+a+\eta) \bigg\}+2J\sinh\eta \, \phi(2a)-\frac{JN\cosh\eta[\cosh^2(2a)-\cosh(2\eta)]}{\sinh^2\eta}.
\eea

Next, we check above results numerically.
Numerical solutions of the BAEs and exact diagonalization of the Hamiltonian (\ref{Ham}) are performed for the case of $2N = 4$ with randomly choosing
of model parameters. The results are listed in Table \ref{roots_real}. We see that the eigenvalues obtained by solving the BAEs
are exactly the same as those obtained by the exact diagonalization of the Hamiltonian (\ref{Ham}). Meanwhile, the expression (\ref{energy_real}) gives the complete spectrum of the system.
\begin{table}[!htbp]
\caption{Numerical solutions of the BAEs (\ref{orBAEs}) for real $\eta$ case, where $J=1$, $\eta=1$, $b=1$, $2N=4$, $n$ indicates the number of the energy levels
and $E_n$ is the corresponding energy.
The energy $E_n$ calculated from the Bethe roots is exactly the same as that from the exact diagonalization of the Hamiltonian (\ref{Ham}). }\label{roots_real}
\small
\begin{center}
\begin{tabular}{|c|c|c|c|c|c|}
\hline $ \lambda_1 $ & $ \lambda_2 $ &$ \lambda_3 $ &$ \lambda_4 $ &$ E_n $ &$ n $ \\ \hline
$-1.0873-1.5708i$ & $-0.5000-0.4634i$ & $-0.5000+0.4634i$ & $0.0873-1.5708i$ & $-5.8897$ &  $1$ \\ \hline
 $-1.7292-1.5708i$ & $-0.5000-1.2559i$ & $-0.5000+1.2559i$ & $0.7292-1.5708i$ & $-5.8897$ &  $1$ \\
 \hline
 $-1.4553+1.0461i$ & $-0.5000-1.1872i$ & $-0.5000+0.8274i$ & $0.4553+1.0461i$ & $-3.9899$ &  $2$ \\
 \hline
 $-1.4553-1.0461i$ & $-0.5000-0.8274i$ & $-0.5000+1.1872i$ & $0.4553-1.0461i$ & $-3.9899$ &  $2$ \\
 \hline
 $-1.7487+1.3618i$ & $-0.5000+0.7332i$ & $-0.5000+1.4005i$ & $0.7487+1.3618i$ & $-3.9899$ &  $2$ \\
 \hline
 $-1.7487-1.3618i$ & $-0.5000-1.4005i$ & $-0.5000-0.7332i$ & $0.7487-1.3618i$ & $-3.9899$ &  $2$ \\
 \hline
 $-1.6037+1.2858i$ & $-0.5000-0.4949i$ & $-0.5000+1.1518i$ & $0.6037+1.2858i$ & $-3.0796$ &  $3$ \\
 \hline
 $-1.6037-1.2858i$ & $-0.5000-1.1518i$ & $-0.5000+0.4949i$ & $0.6037-1.2858i$ & $-3.0796$ &  $3$ \\
 \hline
 $-1.7809+0.6635i$ & $-0.9098+0.3143i$ & $-0.0902+0.3143i$ & $0.7809+0.6635i$ & $3.9899$ &  $4$ \\
 \hline
 $-1.7809-0.6635i$ & $-0.9098-0.3143i$ & $-0.0902-0.3143i$ & $0.7809-0.6635i$ & $3.9899$ &  $4$ \\
 \hline
 $-2.0629-1.2897i$ & $-0.9988-1.1945i$ & $-0.0012-1.1945i$ & $1.0629-1.2897i$ & $3.9899$ &  $4$ \\
 \hline
 $-2.0629+1.2897i$ & $-0.9988+1.1945i$ & $-0.0012+1.1945i$ & $1.0629+1.2897i$ & $3.9899$ &  $4$ \\
 \hline
 $-1.4107+0.0000i$ & $-1.2598-0.0000i$ & $0.2598+0.0000i$ & $0.4107-0.0000i$ & $4.2251$ &  $5$ \\
 \hline
 $-2.0830-1.5708i$ & $-1.0046-1.5708i$ & $0.0046-1.5708i$ & $1.0830-1.5708i$ & $4.2251$ &  $5$ \\ \hline
 $-2.0016-0.9831i$ & $-1.0025-0.8073i$ & $0.0025-0.8073i$ & $1.0016-0.9831i$ & $4.7442$ &  $6$ \\ \hline
 $-2.0016+0.9831i$ & $-1.0025+0.8073i$ & $0.0025+0.8073i$ & $1.0016+0.9831i$ & $4.7442$ &  $6$ \\
 \hline\end{tabular}
 \end{center}\end{table}

%%%%%%%%%%%%%%%%%%%%%%%%%%%%%%%%%%%%%%%%%%%%%%%%%%%%%%%%%%%%%%%
%                                                             %
%               J=1 region                                    %
%                                                             %
%                                                             %
%                                                             %
%%%%%%%%%%%%%%%%%%%%%%%%%%%%%%%%%%%%%%%%%%%%%%%%%%%%%%%%%%%%%%%
\section{Ground state and twisted boundary energy}\label{4}
\setcounter{equation}{0}

In this section, we consider the case of $J=1$. We first analyze the contribution of the third term in the inhomogeneous $T-Q$ relation (\ref{TQ}). For simplicity, we constrain that
$\eta$ is real and $a$ is imaginary. We first introduce following homogeneous $T-Q$ relation
\bea\label{hom_TQ}
\Lambda_h(u)=e^ua(u)\frac{Q_h(u-\eta)}{Q_h(u)}-e^{-u-\eta}d(u)\frac{Q_h(u+\eta)}{Q_h(u)},
\eea
where $Q_h(u)$ is a trigonometric polynomial of the type
\bea
Q_h(u)=\prod_{j=1}^{M}\frac{\sinh(u-\lambda_j)}{\sinh\eta}, \quad M=1,\cdots,2N.
\eea
The period of Bethe roots $\{\lambda_j\}$ is $2\pi$, thus we fix the real part of Bethe roots in the interval $[-\pi, \pi)$.
For convenience, we put $\lambda_j = i u_j/2 - \eta/2$ and $a=ib$. The singularity of Eq.(\ref{hom_TQ}) gives
\bea\label{orBAEs-hom}
&&\hspace{-0.6truecm}e^{iu_j}\left[\frac{\sin\frac{1}{2}(u_j-2b-i\eta)\sin\frac{1}{2}(u_j+2b-i\eta)}
{\sin\frac{1}{2}(u_j-2b+i\eta)\sin\frac{1}{2}(u_j+2b+i\eta)}\right]^N
=\prod^{M}_{l=1}\frac{\sin\frac{1}{2}(u_j-u_l-2\eta i)}{\sin\frac{1}{2}(u_j-u_l+2\eta i)}, \no \\[4pt] &&\quad j=1,\cdots,M.
\eea

Define
\bea
E_h&=& \frac{JN\cosh\eta[\cosh^2(2a)-\cosh(2\eta)]}{\sinh^2\eta} -J\phi^{1-N}(2a)\sinh\eta
\no \\ && \times J\bigg\{ \Lambda_h(a-\eta)\frac{\partial \Lambda_h(u)}{\partial u}\big|_{u=a}+ \Lambda_h(-a-\eta) \frac{\partial \Lambda_h(u)}{\partial u}\big|_{u=-a} \bigg\} \no \\
&=& -\sum_{j=1}^{M}4\pi\sinh\eta \, \phi(2a)[a_1(u_j+2b)+a_1(u_j-2b)]+2\sinh\eta \, \phi(2a) \no \\ && -\frac{N\cosh\eta[\cosh^2(2a)-\cosh(2\eta)]}{\sinh^2\eta},\label{E-homo}
\eea
where the function $a_n(x)$ is given by
\begin{equation}\label{afun-image}
 a_n(x)= \frac{1}{2\pi}\frac{\sinh(n\eta)}{\cosh (n\eta)-\cos x},
\end{equation}
and the Bethe roots $\{u_j\}$ satisfy the BAEs (\ref{orBAEs-hom}).
Taking the logarithm of Eq.(\ref{orBAEs-hom}), we have
\begin{eqnarray}\label{logBAEs-homo}
&& u_j + N [\theta_1(u_j+2b) + \theta_1(u_j-2b)]=2 \pi I_j +\sum^M_{k=1} \theta_2(u_j-u_k), \no \\ && \quad \quad j=1,\ldots,M,
\end{eqnarray}
where the quantum number $\{ I_j \}$ are certain integers (or half odd integers) if $M$ is even (or odd),
\begin{equation}\label{theta}
  \theta_n(x)=2\arctan \frac{\tan (x/2)}{\tan (n\eta/2)}+2\pi \lfloor \frac{x}{2\pi}+
  \frac{1}{2}\rfloor, \no
\end{equation}
and $\lfloor \rfloor$ is the Gauss mark.

From the analysis of Eq.(\ref{logBAEs-homo}) and numerical calculations, we know that the energy $E_{h}$ arrives at its minimum $E^{g}_{h}$ when $M=N$.
Meanwhile, all the Bethe roots $\{u_j\}$ are real and the corresponding quantum numbers are
\begin{equation}\label{GSqnumber}
  I_j=-\frac{N}{2}+1, -\frac{N}{2}+2, \cdots , \frac{N}{2}.
\end{equation}
From Eq.(\ref{GSqnumber}), we see that there is a hole in the real axis which can be put at the boundary $x_0=-\pi$ to minimize the energy.

\begin{figure*}[htbp]
      \centering
    \begin{minipage}[b]{0.49\textwidth}
      \includegraphics[width=1\textwidth]{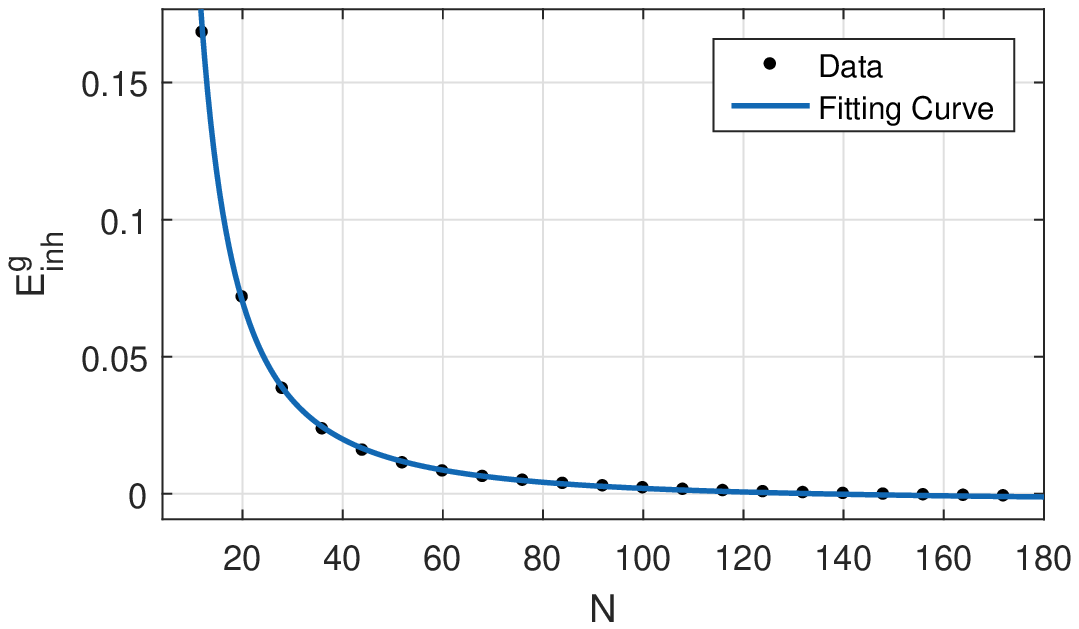}
       \centerline{(a) $a=0.3$}
    \end{minipage}
    \begin{minipage}[b]{0.49\textwidth}
      \includegraphics[width=1\textwidth]{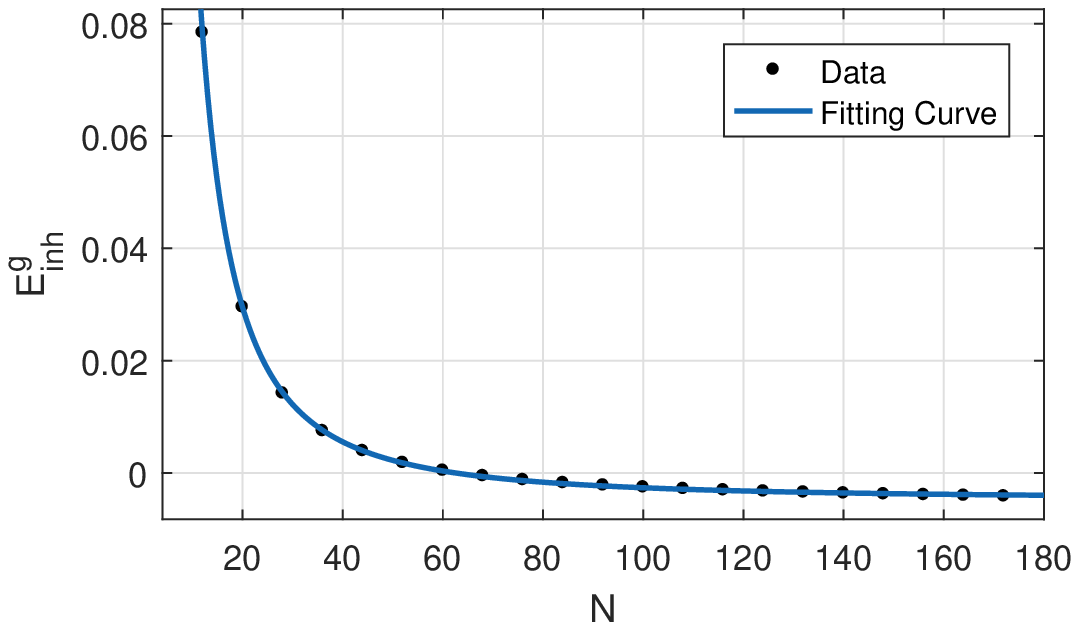}
       \centerline{(b) $a=0.75$}
    \end{minipage}
\caption{The contribution of the inhomogeneous term to the ground state energy with $\eta=2$ and (a) $a=0.3$; (b) $a=0.75$. The data can be fitted as (a)$E^{g}_{inh}=10.96N^{-1.673}-0.0021$; (b)$E^{g}_{inh}=6.237N^{-1.738}-0.0043$. When the system-size tends to infinity, the contribution of the inhomogeneous term can be neglected.}\label{DMRG_12_8_172}
\end{figure*}

Denote the true ground state energy of Hamiltonian (\ref{Ham}) as $E^{g}$. We use the physical
quantity
\bea
E^{g}_{inh}=E^{g}_{h}-E^{g},\label{GSqnumber1}
\eea
to characterize the contribution of the inhomogeneous term in the $T-Q$ relation (\ref{TQ}) at the ground state.
By using the density matrix renormalization group (DMRG) method, we obtain the ground state energy $E^{g}$ of the Hamiltonian (\ref{Ham}). By solving the BAEs (\ref{logBAEs-homo}) and substituting
the values of Bethe roots into the Eq.(\ref{E-homo}), we obtain the energy $E^{g}_{h}$. Substituting $E^{g}$ and $E^{g}_{h}$ into Eq.(\ref{GSqnumber1}), we obtain the values of $E^{g}_{inh}$ and the results are shown as the dots in Fig.\ref{DMRG_12_8_172}. The fitting of the data gives that the contribution of the inhomogeneous term tends to zero when the system-size tends to infinity. Then, we conclude that
the Eq.(\ref{E-homo}) is a suitable approximation of the energy of the system (\ref{Ham}) in the thermodynamic limit. In the following, we use $E^{g}_{h}$
to quantity the ground state energy of the model (\ref{Ham}).

Define
\begin{equation}\label{cou-fun}
 Z(u)= \frac{1}{4\pi} \bigg[  \frac{u}{N}+ \theta_1(u+2b) + \theta_1(u-2b) - \frac{1}{N}\sum^M_{k=1} \theta_2(u-u_k) \bigg] .
\end{equation}
Obviously, $Z(u_j ) = I_j/2N$ corresponds to the Eq.(\ref{logBAEs-homo}).
In the thermodynamic limit, $N \rightarrow \infty$, $M \rightarrow \infty$ and $N /M$  finite, the counting function $Z(u)$ becomes a continue function of $u$.
Taking the derivative of Eq.(\ref{cou-fun}) with respect to $u$, we obtain
\begin{eqnarray}\label{den-1}
\frac{d Z(u)}{d u} &=& \frac{1}{4\pi N} + \frac{1}{2}  \left[ a_1(u+2b) +a_1(u-2b) \right] - \int^\pi_{-\pi} a_2(u-\lambda) \rho(\lambda) d\lambda \nonumber  \\[4pt]
&\equiv &  \rho(u)+\rho^h(u),
\end{eqnarray}
where $\rho(x)$ and $\rho^h(x)$ are the densities of particles and holes, respectively.
The Fourier transformation of $a_n(x)$ is
\begin{equation}\label{afun-FT-image}
  \tilde{a}_n(\o)\!=\!\frac{1}{2\pi}\int_{-\pi}^{\pi}\!e^{i\o x}\frac{\sinh(n\eta)}{\cosh (n\eta)-\cos x}dx\!=\!e^{-n\eta|\o|}.\!
\end{equation}
At the ground state, there is a hole at the point of $x_0=-\pi$. Thus the density of holes reads
\bea
\rho^h_g(x)=\frac{\delta(x-x_0)}{2N},
\eea
and the corresponding Fourier transformation is
\bea
\tilde{\rho}^h_g(w)=\frac{e^{-i\o x_0}}{2N}.
\eea
Taking the Fourier transformation of Eq.(\ref{den-1}), we obtain
\bea\label{rho_g}
\tilde{\rho}_g(\o ) = -\frac{e^{i\o \pi}}{2N(1+e^{-2\eta|\o|})} + \frac{\delta_{\o,0}}{2N(1+e^{-2\eta|\o|})}+\frac{\cos(2b\, \o )}{2\cosh(\eta\,\o)}.
\eea
The ground state energy of model (\ref{Ham}) is
\bea\label{GSE}
E^g&=&-4N\sinh\eta\,\phi(2bi)\sum_{\o=-\infty}^{\infty}\frac{e^{-\eta|\o|}\cos^2(2b\o)}{\cosh(\eta \o)}-\frac{N\cosh\eta[\cos^2(2b)-\cosh(2\eta)]}{\sinh^2\eta} \no \\ && +2\sinh\eta\,\phi(2bi)\sum_{\o=-\infty}^{\infty}\frac{e^{i\pi\o}\cos(2b\o)}{\cosh(\eta \o)}.
\eea

The twisted boundary energy is defined as
\bea\label{TBE}
E_b=E^g-E^g_p,
\eea
where $E^g_p$ is the ground state energy for the Hamiltonian (\ref{Ham}) with periodic boundary condition \cite{Qia19}
\bea\label{GSE-P}
E^g_{p}=-4N\sinh\eta\,\phi(2bi)\sum_{\o=-\infty}^{\infty}\frac{e^{-\eta|\o|}\cos^2(2b\o)}{\cosh(\eta \o)}-\frac{N\cosh\eta[\cos^2(2b)-\cosh(2\eta)]}{\sinh^2\eta}.
\eea
Substitute Eqs.(\ref{GSE}) and (\ref{GSE-P}) into (\ref{TBE}), we obtain the value of twisted boundary energy as
\bea
E_b=2\sinh\eta\,\phi(2bi)\sum_{\o=-\infty}^{\infty}\frac{e^{i\pi\o}\cos(2b\o)}{\cosh(\eta \o)}.
\eea

%%%%%%%%%%%%%%%%%%%%%%%%%%%%%%%%%%%%%%%%%%%%%%%%%%%%%%%%%%%%%%%
%                                                             %
%               J=-1 region                                   %
%                                                             %
%                                                             %
%                                                             %
%%%%%%%%%%%%%%%%%%%%%%%%%%%%%%%%%%%%%%%%%%%%%%%%%%%%%%%%%%%%%%%

\section{Bethe roots of inhomogeneous BAEs}\label{5}

\setcounter{equation}{0}

\begin{figure}[!htbp]
\centering
\includegraphics[height=7cm,width=8cm]{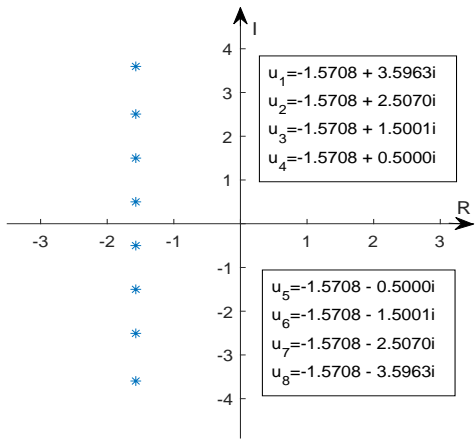}
\caption{The distributions of Bethe roots at the ground state, where $\eta=1, a=0.3i$ and $2N=8$. The string structure can be seen very clearly from the data.}\label{N-strings}
\end{figure}

In this section, we consider the case $J=-1$.
In this case, the ground state spin configuration and the solution of Bethe roots in
BAEs (\ref{orBAEs}) are different from those with $J=1$. Again, $\eta$ is set as real and $a$ is set as imaginary.
For convenience, put $\lambda_j=i u_j-\eta/2$. The BAEs (\ref{orBAEs}) become
\bea\label{BAEs2}
\frac{e^{iu_j}\prod_{l=1}^{2N}\sin(u_j-u_l+i\eta)}{\sin^N(u_j+b+\frac{1}{2}i\eta) \sin^N(u_j-b+\frac{1}{2}i\eta)}&=&
\frac{e^{-iu_j}\prod_{l=1}^{2N}\sin(u_j-u_l-i\eta)}{\sin^N(u_j+b-\frac{1}{2}i\eta) \sin^N(u_j-b-\frac{1}{2}i\eta)} \no \\ [5pt] &&
\hspace{-7truecm}+2i\,e^{-N\eta}\sin\big(u_j-\sum_{l=1}^{2N}{u_l}\big),
\quad \quad j=1,\cdots,2N.
\eea
By careful analysis and numerical check, we find the Bethe roots in Eq.(\ref{BAEs2}) form the $2N$-strings at the ground state
\bea\label{N-string}
u_j=x_0+\big(\frac{N+1}{2}-j\big) i \eta +o(2N), \quad j=1,\cdots,2N,
\eea
where $x_0$ is real and $o(2N)$ stands for a small correction which is related with $2N$ and $i$ is the imaginary unit.
The numerical results of Bethe roots at the ground state with $2N=8$ is shown in Fig.\ref{N-strings}. From which, the string structure of Bethe roots can been seen very clearly.
Substituting the string hypothesis (\ref{N-string}) into the energy expression (\ref{energy_real}) and neglecting the small correction, we obtain the energy for the $2N$-string states
\bea\label{E-N-string}
E_{s}&=&-2\sinh\eta\,\phi(2a)\bigg[ \frac{\sinh(2N\eta)}{\cos(2x_0+2b)-\cosh(2N\eta)} +\frac{\sinh(2N\eta)}{\cos(2x_0-2b)-\cosh(2N\eta)} \bigg] \no \\ [5pt] && -2\sinh\eta \,\phi(2a)+ \frac{N\cosh\eta[\cosh^2(2a)-\cosh(2\eta)]}{\sinh^2\eta}.
\eea
When system size $2N$ is very large, the energy for the $2N$-string (\ref{E-N-string}) turns to
\bea\label{E_stringGE}
E_{s}=4\sinh\eta\,\phi(2a)\tanh(2N\eta) -2\sinh\eta \,\phi(2a)+ \frac{N\cosh\eta[\cosh^2(2a)-\cosh(2\eta)]}{\sinh^2\eta}.
\eea
We see that the $2N$-string energy (\ref{E_stringGE}) is irrelevant with the position $x_0$ of string.

\begin{figure}[!htbp]
\begin{minipage}{0.30\linewidth}
  \leftline{\includegraphics[width=8.5cm]{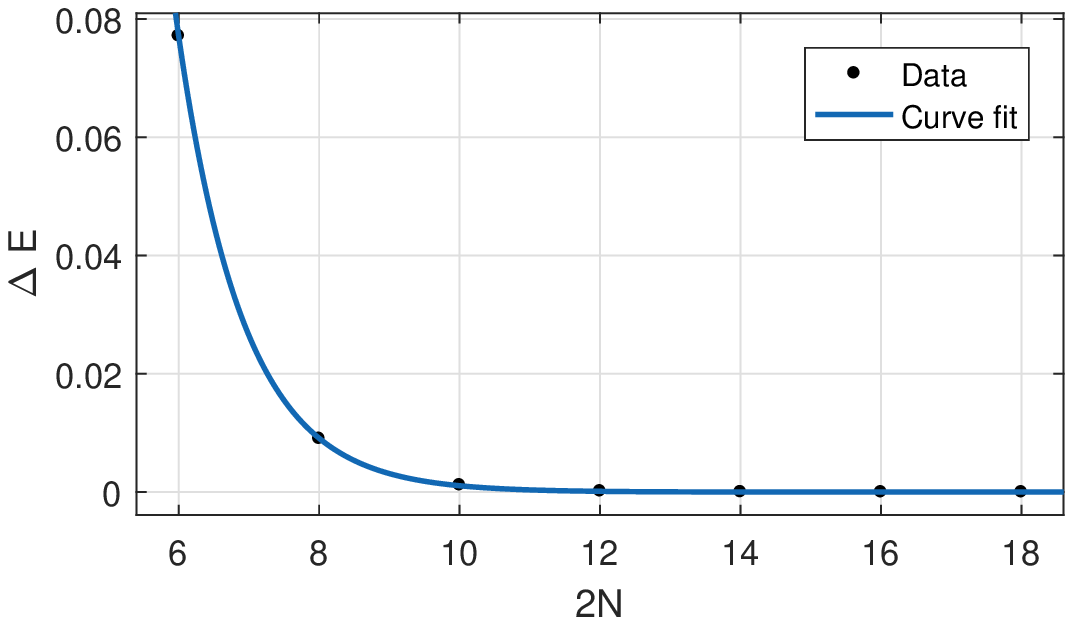}}%_67.56e(-1.041x)
  \rightline{(a) $\eta=1, a=0.3i$ \hspace{-1.3truecm}}
\end{minipage}
\hfill
\begin{minipage}{0.30\linewidth}
  \rightline{\includegraphics[width=8.5cm]{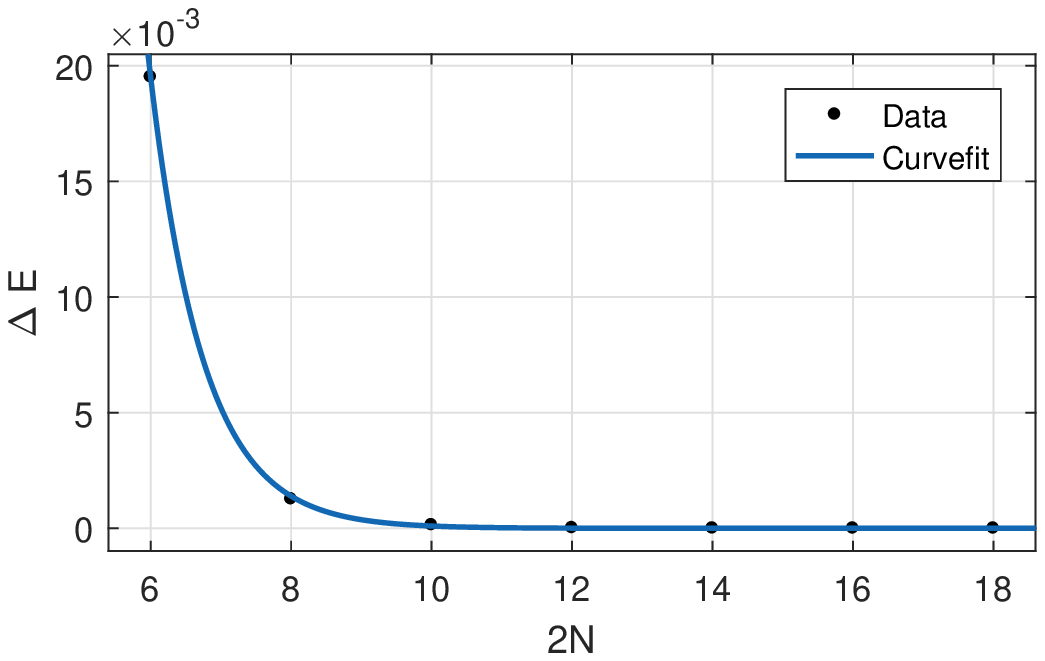}}%_427.4e(-1.999x)
  \leftline{\hspace{-1.3truecm}(b) $\eta=1, a=0.75i$ }
\end{minipage}
\caption{The energy difference $\Delta E = E_{s}^{}-E^{ED}$, where
$E_{s}^{}$ is the ground state energy calculated by the string solution (\ref{E_stringGE}) and $E^{ED}$ is the ground state energy calculated by the exact diagonalization.
The $\Delta E$ is exponentially decreasing with the increasing system-size $2N$ and the data can be fitted as (a) $\Delta E = 46.9e^{-1.068*(2N)}$ (b) $\Delta E = 53.7e^{-1.32*(2N)}$. Thus, the energy difference $\Delta E$ will be zero in the thermodynamic limit.}\label{GSE_plot}
\end{figure}

In order to show the correction of Eq.(\ref{E_stringGE}), we calculate the ground state energy of the system (\ref{Ham}) by exactly diagonalizing the Hamiltonian up to $2N=18$
and compare the results with those obtained from Eq.(\ref{E_stringGE}) in Fig.\ref{GSE_plot}. From it, we see that the energy difference $\Delta E$, which comes from the finite-size effect of the string (\ref{N-string}),
exponentially tends to zero with the increasing $2N$. The data satisfy the scaling law, $\Delta E \propto e^{ \alpha 2N }, \alpha < 0$. Thus in the thermodynamic limit,
the expression (\ref{E_stringGE}) gives the exact value of the ground state energy.

Now, we calculate the twisted boundary energy. The ground state for the Hamiltonian (\ref{Ham}) with periodic boundary condition is the ferromagnetic state if $J=-1$. It is easy to obtain the ground state energy as
\bea\label{E_periodGE}
E^p=-2N\cosh\eta+\frac{N\cosh\eta\sinh^2(2a)}{\sinh^2\eta}.
\eea
From the definition of twisted boundary energy
\bea \label{boundary1}
E_{t}=E_{s}-E^p,
\eea
and substituting Eqs. (\ref{E_stringGE}) and (\ref{E_periodGE}) into (\ref{boundary1}), we obtain
\bea \label{boundary111}
E_{t}=4\,\phi(2a)\sinh\eta\,\tanh(2N\eta) -2\sinh\eta \,\phi(2a).
\eea
In the thermodynamic limit, the twisted boundary energy arrives at
\bea
E_{t}=2\sinh\eta\,\phi(2a). \label{boundary2}
\eea

%%%%%%%%%%%%%%%%%%%%%%%%%%%%%%%%%%%%%%%%%%%%%%%%%%%%%%%%%%%%%%%
%                                                             %
%               Conclusion                                    %
%                                                             %
%                                                             %
%                                                             %
%%%%%%%%%%%%%%%%%%%%%%%%%%%%%%%%%%%%%%%%%%%%%%%%%%%%%%%%%%%%%%%

\section{Conclusion}\label{6}

In this paper, we study an integrable anisotropic $J_1-J_2$ spin chain with antiperiodic boundary condition.
By means of the off-diagonal Bethe Ansatz, we obtain the exact solution of the system.
We show that the contribution of inhomogeneous term in the $T-Q$ relation can be neglected when the system-size tends to infinity. Based on it, we discuss the ground state energy and the twisted boundary energy.
We find the string structure of Bethe roots at the ground state if the coupling constant $J=-1$ although the corresponding Bethe Ansatz equations are the inhomogeneous ones.

\section*{Acknowledgments}

We would like to thank Prof. Y. Wang for his valuable discussions and continuous encouragements.
The financial supports from the National Program
for Basic Research of MOST (Grant Nos. 2016YFA0300600 and
2016YFA0302104), the National Natural Science Foundation of China
(Grant Nos. 11934015, 11434013, 11425522, 11547045, 11774397, 11775178 and 11775177), the Major Basic Research Program of Natural Science of Shaanxi Province
(Grant Nos. 2017KCT-12, 2017ZDJC-32), Australian Research Council (Grant No. DP 190101529), the Strategic Priority Research Program of the Chinese
Academy of Sciences and the Double First-Class University Construction Project of Northwest University are gratefully acknowledged.

%%%%%%%%%%%%%%%%%%%%%%%%%%%%%%%%%%%%%%%%%%%%%%%%%%%%%%%%%%%%%%%
%                                                             %
%   Reference                                                 %
%                                                             %
%                                                             %
%                                                             %
%%%%%%%%%%%%%%%%%%%%%%%%%%%%%%%%%%%%%%%%%%%%%%%%%%%%%%%%%%%%%%%
%\fussy
%\bibliographystyle{unsrt}
%\bibliography{ref1}

\end{document}